\documentstyle[11pt,newpasp,twoside,epsf]{article}
\def\edcomment#1{\iffalse\marginpar{\raggedright\sl#1\/}\else\relax\fi}
\reversemarginpar

\newcommand{\nhi}{\mbox{$N_{\rm HI}$}} 
\newcommand{\hi}{H~{\sc i}}

\newcommand{\mhi}{\mbox{$M_{\rm HI}$}}

\newcommand{\msol}{\mbox{${\rm M}_\odot$}}

\newcommand{\fn}{\mbox{$f(N)$}}
\newcommand{\dla}{DL$\alpha$}
\newcommand{\icmsq}{\mbox{$\rm cm^{-2}$}}

\newcommand{\dndz}{\mbox{$dN/dz$}}
\newcommand{\hubble}{\mbox{$\rm km\, s^{-1}\, Mpc^{-1}$}}

\begin{document}
\title{Reconciling Damped Ly-$\alpha$ Statistics
and 21cm Studies at $z=0$}
\author{Martin Zwaan}
\affil{         School of Physics,
                University of Melbourne,
                Victoria 3010,
                Australia}
\author{Frank Briggs}
\affil{Kapteyn Astronomical Institute, PO Box 800, 9700 AV, Groningen,
    The Netherlands}
\author{Marc Verheijen}
\affil{Department of Astronomy, University of Wisconsin, 475 North
    Charter Street, Madison, WI 53706}


\begin{abstract} Blind 21cm surveys in the local universe have shown
that the local \hi\ mass density, $\Omega_{\rm HI}$, is dominated by
luminous, high surface brightness, spiral galaxies.  On the other hand,
surveys for host galaxies of damped Ly-$\alpha$ systems have not always
been successful in finding bright spiral galaxies.  From an analysis of
21cm aperture synthesis maps of nearby galaxies we show that this
apparent contradiction can be resolved by realizing that the \hi\ {\em
mass} density is dominated by $L^*$ galaxies, but the \hi\ {\em cross
section} near the \dla\ threshold is more evenly distributed over
galaxies with a large range in luminosity, gas mass, and surface
brightness.  The distributions of column densities and impact parameters
of optically identified {\it and} non-identified \dla\ host galaxies in
the literature and the \hi\ maps are qualitatively in agreement.  Due to
poor number statistics of low redshift \dla\ systems, there is no firm
indication that the redshift number count of low redshift \dla\ systems
is inconsistent with that calculated from the nearby galaxy population. 
\end{abstract}

\section{Introduction}
 Blind 21cm surveys and studies of damped Ly-$\alpha$ (\dla) systems use
completely unrelated observational techniques but they both can be employed to
address similar scientific questions: what is the cosmological mass
density of \hi\ ($\Omega_{\rm HI}$), what sort of galaxies contribute
most to $\Omega_{\rm HI}$, and how does $\Omega_{\rm HI}$ at $z=0$
relate to that at high $z$? The main advantage of the 21cm surveys is
that the total \hi\ mass of the detected cloud or galaxy can be readily
derived, and after follow-up with synthesis instruments a 3-dimensional
data cube can be obtained from which a detailed map of the \hi\
distribution and a velocity field can be extracted.  Unfortunately, the
neutral hydrogen 21cm emission line is weak, and blind 21cm surveys are
limited to the very local ($z<0.1$) universe.  In contrast, distance is
not a limiting factor in the identification of \dla\ systems, since their
detection depends on the brightness of the background source against
which the absorber is observed.  However, the nature of these systems
is more difficult to determine
because the absorption spectrum gives information along one
very thin sight-line through the
absorber, and the total \hi\ mass of individual
absorption systems can not be determined.
Furthermore,
because \dla\ systems are often found at high redshift, determining their
optical properties requires extensive, time-consuming imaging programs. 

One of the major challenges of 21cm and QSO absorption line research is
to reconcile the results from blind \hi\ surveys and \dla\ surveys at
low redshift.  The aim of this paper is to briefly review the results
from both fields, and test whether the results can be brought in
agreement.

\section{Results from 21cm and \dla\ surveys}
 The most extensive blind 21cm surveys to date are the Arecibo \hi\
Strip Survey (AHISS, Zwaan et al.  1997), the Arecibo Dual Beam Survey
(ADBS, Rosenberg \& Schneider 2000), and the HI Parkes Sky Survey
(HIPASS, Staveley-Smith et al.  1996, Kilborn 2001).  
The main results from the AHISS,
relevant to the present discussion, are that ({\em i\/}) the local \hi\
mass density is dominated by \hi\ rich, high surface brightness,
luminous spiral galaxies (Zwaan et al.  1997, Zwaan, Briggs \&
Sprayberry 2001).  Galaxies with \hi\ masses below $10^{8.5}\msol$ make up
only about 20\% of the neutral gas density, and low surface brightness
(LSB) galaxies with central $B$-band surface brightness $>23.0\, \rm mag
arcsec^{-2}$ contribute a similar fraction, and ({\em ii\/}) where there is \hi\
there are stars.  No convincing instance of star-less extragalactic \hi\
cloud, not gravitationally bound to a parent galaxy, has yet been
identified.  The results from the ADBS and preliminary results from
HIPASS are in agreement with these conclusions.

Thus, if most of the \hi\ in the local universe is in these conspicuous
galaxies, and at $z>0$ \dla\ systems contain most of the \hi, this would
naively lead to the conclusion that \dla\ host galaxies
should be predominantly bright spiral galaxies and thus
easily identifiable.
However, this does not appear to be the case.
The tally so far is that
optical imaging of $z<1$ \dla\ systems has turned up
5 LSB irregulars, 3 compact dwarfs, and $\pm 5$ `normal' spiral galaxies
(Le Brun et al. 2000, Nestor et al. 2001).
In $\pm 2$ cases (depending on which reference to choose) a
host galaxy has not been identified.  H$\alpha$ emission has been
searched for in only one \dla\ system by Bouch{\'e} et al.  (2000). 
They did not detect H$\alpha$ emission in their deep HST imaging of a
$z=0.656$ \dla\ system, which allows them to put an upper limit of
$1.9~M_{\odot}/\rm yr$ on the global star formation rate in \dla\
system, which is lower than the typical value measured in $L^*$ field
spiral galaxies.  A third method of studying properties of \dla\ systems is to
search for 21cm emission in order to measure the total cold gas content
of the \dla\ system.  The problem with these surveys is that the 21cm
\hi\ emission line is extremely weak so that with present technology
these surveys are limited to the very local ($z<0.2$) Universe.  Two
attempts to detect \hi\ in emission in \dla\ systems have been made by
Kanekar et al. (2001) and by Lane et al.  (2000).  Both searches
resulted in non-detections, which allowed the authors to put 
upper limits to the \hi\ masses of approximately $\mhi=3
\times 10^9 \msol$ ($H_0=65\, \hubble$), which is one third of the \hi\ mass of
an $L^*$ galaxy.  A reverse strategy was followed by Bowen et al. 
(2001), who identified a \dla\ absorber in a nearby ($z=0.01$) very
gas-rich LSB galaxy. 

The conclusion from these surveys is that low redshift \dla\ host
galaxies display a wide range in galaxy parameters.  This is a
conclusion very similar to that from Nestor et al.  (2001) who discuss
the results of an optical and NIR imaging survey of five QSO fields with
identified \dla\ systems with $z<0.53$.  Obviously, it are not just the
bright, high surface brightness spiral galaxies that dominate the cross
section for $\nhi > 2\times 10^{20} \icmsq$.  Is this a contradiction to
the low redshift 21cm emission line survey results? To answer this
question it is essential to study the $\nhi$ distribution in nearby
galaxies in more detail. 

\section{Measuring the local \hi\ column density distribution function}
 Determination of the local \hi\ column density distribution function
requires 21cm maps of a fair sample of galaxies that is
representative of the nearby
galaxy population.  Few such samples exist, because
traditionally  radio astronomers have chosen to study
galaxies with extended \hi\
disks, so that the rotation curve can be measured out to large radii, or
galaxies with peculiarities or participating in interactions.
The currently available data from \hi\
selected galaxy samples are not appropriate
because the spatial resolution of
the survey instruments (Arecibo or Parkes) is too coarse to make
detailed \hi\ maps.  21cm follow-up on some of these samples (AHISS, ADBS)
has been done with the VLA in the most compact (D) configuration, which only
barely resolves the galaxies.  Using these low resolution
\hi\ maps  at a particular limiting column density contour
leads to an overestimation of the effective cross-section
for \dla\ absorption.
The exception is the ongoing
work of Ryan-Weber et al.  (2001), who use the Australian Telescope Compact
Array (ATCA) to study a large sample of nearby galaxies taken from the
HIPASS catalogue.  Eventually, this sample will yield a valuable
measurement of the local \hi\ column density distribution function. 

In Zwaan, Verheijen \& Briggs (1999) we presented a measurement of the
$z=0$ \hi\ column density distribution function [\fn], derived from deep
WSRT 21cm maps of a volume-limited sample of galaxies in the Ursa Major
(UMa) cluster (Verheijen \& Sancisi 2001).  We argued
that this sample of
galaxies is representative of the field because ({\em i\/}) the collapse
time of the Ursa Major cluster is close to a Hubble time which implies
that few interactions have taken place, ({\em ii\/}) the \hi\ maps
show no signs of \hi\ depletion, and ({\em iii\/}) the morphological mix
of the galaxy population in the Ursa Major cluster is similar to that in
the field.  The sample is very suitable for calculating \fn\ because all
galaxies are at approximately the same distance, which yield a uniform
spatial resolution of $\sim 1$~kpc.  The sample consists of 49 galaxies,
and is complete to an absolute magnitude $M_B=-16.9$, roughly
twice the luminosity of the Small Magellanic Cloud.

The UMa \fn, presented in Figure 2 of Zwaan et al.  (1999), was a
considerable improvement over the function previously derived by Rao \&
Briggs (1993), who used Gaussian fits to the radial \hi\ distribution of
27 galaxies with large optical diameters, as measured with the Arecibo
Telescope.  The UMa analysis was based on much higher spatial resolution
observations ($\sim 15''$ vs.  $3'$), the sample contained more galaxies,
and no modelling was applied.  The UMa \fn\ follows a $\nhi^{-1}$
distribution for \nhi\ $<10^{21} \icmsq$, and a $\nhi^{-3}$ distribution
for \nhi\ $>10^{21} \icmsq$.  

\section{The distribution of cross-sectional area}
 Here we re-analyze the Ursa Major sample to make a more specific
comparison between the available \dla\ data and the information from
nearby galaxy population. 
In Figure~1 we show the `volume density of cross-sectional area' for
different column density cut-offs.  This cross-sectional area per volume
is calculated by multiplying $A(M_B)\times \Phi(M_B)$, where $A(M_B)$ is
the area of the \hi\ above a certain column density limit, as a function
of absolute magnitude $M_B$, and $\Phi(M_B)$ is the galaxy luminosity function. 
The result of this product is $dN/dz \times H_0/c$, where \dndz, the
number of absorbers per unit redshift, is a familiar quantity in QSO
absorption line studies.  To account for the fact that the sample
excludes galaxies with inclinations $i<45^{\circ}$, we have de-projected
the \hi\ maps, and recalculated the area function $A(M_B)$ for the
sample as if it were randomly oriented with inclination $i<45^{\circ}$.
These values have been incorporated in the total value of $A(M_B)$.

\begin{figure}
\epsfxsize=12.5cm \epsfbox[30 470 530 700]{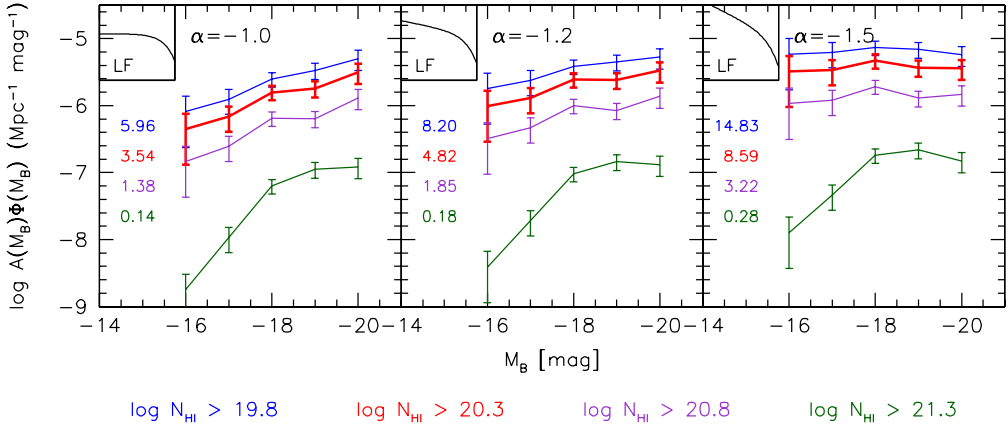} 
\caption{Cross-sectional area of HI in the UMa galaxy sample as a function
of $B$-band absolute magnitude. The different colors correspond to
different column density cut-offs, where the fat, red line is for column
densities $\log \nhi>20.3$, which is the limit above which an HI
absorbing system is called a damped Ly-$\alpha$ system. The three panels
show the effect of using different assumed shapes of the galaxy luminosity
function. The numbers show the values of $100\times dN/dz$, where
$dN/dz$ is integrated over the the whole range in $M_B$ over which the
function is evaluated. Errorbars are calculated using Poisson statistics.
\label{ATheta_MB.fig}
}
\end{figure}                   

To illustrate the effects of different shapes of the luminosity
function, we show the distribution of $A(M_B)\times \Phi(M_B)$ for three
different values of the faint-end slope of the Schechter function:
$\alpha=-1.0$, $\alpha=-1.2$ and $\alpha=-1.5$, as indicated by the
miniature luminosity functions in the upper left corner of each panel. 
The value of $\alpha=-1.2$ is closest to the recently published results
from the 2dF survey (Folkes et al.  1999) and the Sloan Digital Sky
Survey (Blanton et al.  2001).

The main point here is that the distribution of cross section for $\log
\nhi> 20.3$ is much less peaked around $L^*$ galaxies than the
distribution of \hi\ mass density (Figure 3 from Zwaan, Briggs, \&
Sprayberry 2001).  Indeed, the \hi\ cross section in galaxies with
$M_B=-16$ is only slightly lower than that in galaxies with $M_B=-20$.  The
consequence of this observation is that if an \hi\ column density
$>10^{20.3}\,\icmsq$ is encountered somewhere in the local universe, the
probability that this gas is associated with a $L^*/40$ galaxy is almost
as high as for association with a $L^*$ galaxy.  On the other
hand, the highest column densities ($>10^{21.3}\,\icmsq$) are
largely associated
with the most luminous galaxies.  Small cross sections of very high
column density gas can still account for high \hi\ masses, which
explains why the total \hi\ density is dominated by $L^*$ galaxies. 

\begin{figure}
\epsfxsize=12.5cm \epsfbox[30 470 530 700]{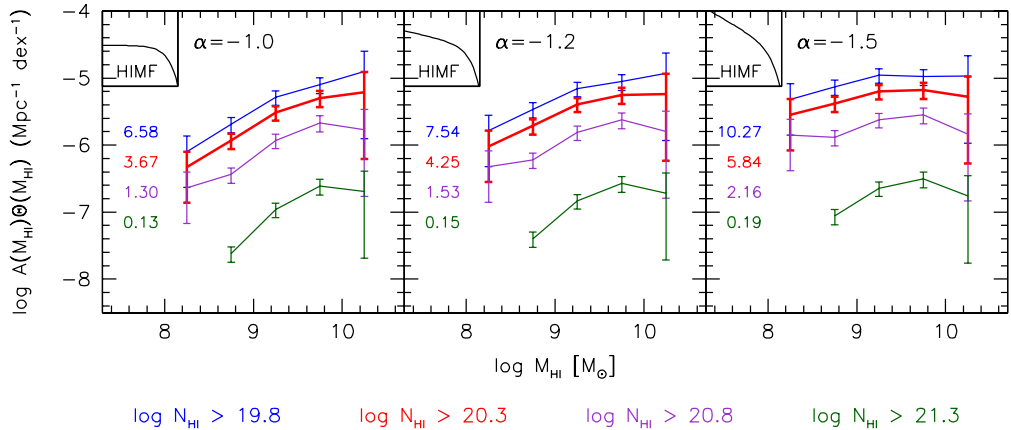} 
\caption{Cross-sectional area of HI in the UMa galaxy sample as a function
of \hi\ mass. (Curves are labeled as described in Fig.~1.)
The \hi\ mass function used for the normalisations are illustrated in
the upper left corner of each panel. 
\label{ATheta_MHI.fig}
}
\end{figure}                   

A similar plot can be constructed for the cross-sectional area as a
function of \hi\ mass.  This is shown in Figure~2, where we plot
$A(M_B)\times \Theta(M_{\rm HI})$, where $\Theta(M_{\rm HI})$ is the
\hi\ mass function.  Again, the results for three different values of
the faint-end slope are shown.  The \hi\ mass fuction based on the
Arecibo \hi\ Strip Survey (AHISS) discussed by Zwaan et al.  (1997), has
a slope $\alpha=-1.2$, corresponding to the middle panel.  Steeper \hi\
mass functions have  been advocated by Rosenberg \& Schneider
(2001) on the basis of the ADB Survey.

The dependence of interception cross section on HI mass of the
host in Fig.~2 shows a similar behavior to that of optical luminosity
(as in Fig.~1), although less
pronounced.  Most of the cross section of \dla\ column densities is
associated with large \hi\ mass galaxies, but the contribution from
lower mass galaxies is not negligible: galaxies with $\mhi<10^9\msol$
contribute $\sim 30\%$ to the cross section.

Finally, Figure~3 shows the cross-sectional area as a function of
$K$-band central surface brightness.  Weights to the individual galaxies
have been calculated using optical luminosity functions, similar to
those in Figure~1.  This is a more robust method of determining the
normalisation than using the central surface brightness distribution
function (the galaxy density as a function of central surface brightness),
which has not been measured to great accuracy.  Concentrating
on the central panel, it appears that galaxies two magnitudes dimmer that
the canonical `Freeman-value' (Freeman 1970) contain a comparable 
amount of cross section at \dla\ column densities to those with
the Freeman central surface brightness.  Again, the distribution
function for the highest column densities is more clearly peaked around
high surface brightness galaxies, which explains the dominance of these
galaxies in $\Omega_{\rm HI}$. 

\begin{figure}
\epsfxsize=12.5cm \epsfbox[30 470 530 700]{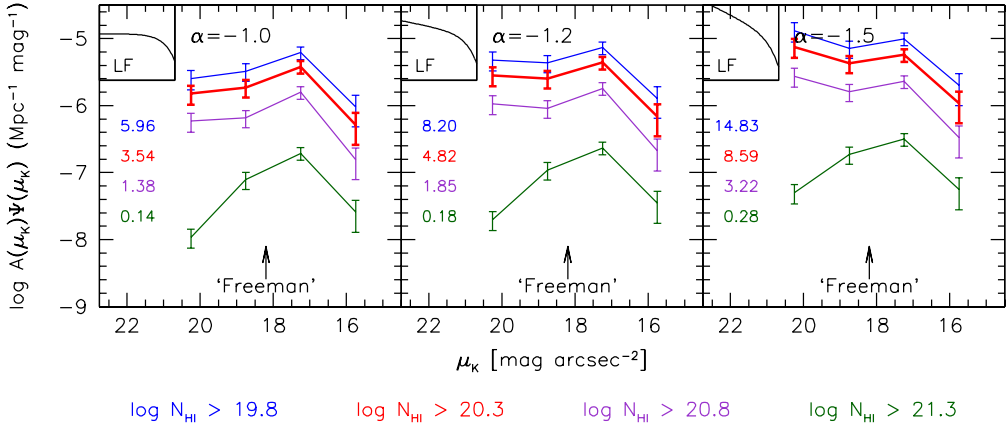} 
\caption{Same as Figure~1, but now as a function of 
$K$-band central surface brightness. The same luminosity function as in
Figure~1 have been used to calculate normalisations. 
The arrow with `Freeman' indicates the $K$-band equivalent of the
`Freeman-value', the mean central surface brightness of high surface
brightness spiral galaxies. 
\label{ATheta_sb.fig}
}
\end{figure}                   

The conclusion from these three figures is that the probability
distribution of cross-sectional area of \dla\ column densities as a
function of various galaxy parameters is much less peaked at $L^*$-type
properties than the \hi\ density.  This explains why surveys for \dla\
host galaxies turn up a population of galaxies with a large spread in
luminosity, morphological type, and surface brightness.  Rosenberg \&
Schneider, and Turnshek et al.  have presented similar plots of cross
section vs.  luminosity at this meeting. 

\section{Redshift number density}
 The above analysis shows that 21cm surveys and \dla\ results are
in qualitative agreement.  The remaining question is whether \dndz,
the number of absorbers per unit redshift, is quantitatively
consistent between the two techniques.

Rao \& Turnshek (2000) recently undertook a HST survey of low redshift
($z<1.65$) Mg{\sc ii} selected systems, with the aim of measuring the
incidence of \dla\ systems.  The underlying assumption that makes this
statistical approach work, relies on the empirical fact that 
all \dla\ systems
are Mg{\sc ii} absorbers.
Rao \&
Turnshek conclude that \dndz\ shows no indication of evolution over the
redshift range $0.5<z<4$, but the $z=0$ point measured from 21cm surveys
is significantly lower.  However, Briggs (2001) argued that the Mg{\sc ii}
statistics used to bootstrap the \dla\ \dndz\ values, should be limited
to those systems with $W_0^{\lambda2796}>1.0\rm\AA$, since the Mg{\sc
ii}-selected \dla\ systems are drawn almost entirely from the large equivalent
width Mg{\sc ii} sub-sample.  Because Steidel \&
Sargent (1992) found that at low redshift these large
$W_0^{\lambda2796}$ systems evolve significantly faster than the low
$W_0^{\lambda2796}$ systems, the lowest $z$ Rao \& Turnshek
values of \dndz\ would
be lower, which could lessen the discrepancy with the $z=0$ point
substantially.
Interestingly, Churchill (2001) performed an unbiased survey
for low-$z$ Mg{\sc ii} systems and used the same statistics for Mg{\sc ii} systems
as Rao \& Turnshek did, and found that \dndz\ for \dla\ absorption is
$0.08^{+0.09}_{-0.05}$ at $\langle z \rangle =0.05$.  This value agrees very well with a
no-evolution extrapolation of the Rao \& Turnshek results. 
Unfortunately, small number statistics (Churchill's result is based on
four systems) causes large uncertainties on the derived values, which
makes the interpretation very difficult.  The $\Omega_{\rm HI}$
studies that depend on intermediary Mg{\sc ii} lines contrast sharply
with lower values from the more traditional \dla\ survey methods
(Jannuzi et al. 1998, Lanzetta et al. 1995), although they also suffer from
small number statistics.

Using the UMa sample and the \hi\ mass function from Zwaan et al. 
(1997) with a faint-end slope of $\alpha=-1.2$, we find for \dla\ column
densities a value of $\dndz(z=0)=0.042\pm 0.015$.  The quoted
uncertainty on this value is based on Poisson statistics, and does not
take into account the uncertainties on the \hi\ mass function
parameters.  If we use the isophotal $B$-band galaxy luminosity function
parameters from Blanton et al.  (2001), and account for the galaxies in
the UMa sample not detected in \hi, we arrive at the same value for
\dndz.  This value is higher than the one found by Rao \& Briggs (1993)
($\dndz=0.015$), probably because these authors limited their
calculation to large, optically bright galaxies, whereas our calculation
is based on a volume-limited galaxy sample, which also includes dwarf
and LSB systems. 

At present it is unclear whether the \dla\ redshift number densities at
$z\sim~0$ derived from 21cm emission maps and Mg{\sc ii} selected systems
agree.  The errorbars on the Churchill value are sufficiently large to
make them in agreement with our estimate (less than $1\sigma$
difference).  Obviously, much better statistics are required to appraise
the results from the different methods.  Rosenberg \& Schneider (2001)
use their VLA D-array imaging of \hi\ selected galaxies to find
$\dndz=0.07 \pm 0.01$.  There are two reasons why this value is higher
than the one we find.  Firstly, the \hi\ mass function that these
authors use has a faint-end slope of $\alpha=-1.5$, whereas we use
$\alpha=-1.2$.  This steeper \hi\ mass function causes low mass galaxies
to contribute more cross section.  Secondly, their analysis is based on
lower resolution data of galaxies at larger distances.  This might lead
to an overestimation of the cross sections. 

\section{Comparing properties of \dla\ systems and nearby galaxies}
 In Figure~4 we show the 2-dimensional probability density distribution
of cross section in the $(\nhi,R)$ plane, where $R$ is distance from the
centre of the galaxy.  We have used the UMa sample to calculate the
cross-sectional area contributed by each element $d\nhi dR$, on a fine
grid in the $(\nhi,R)$ plane.  To increase the signal-to-noise in the
figure, we smoothed this probability distribution with a Gaussian filter
with $\sigma=0.1$ in the $\log \nhi$ direction and $\sigma=1$ in the $R$
direction.  An \hi\ mass function with a faint-end slope $\alpha=-1.2$
was again used to calculate the weights of the individual galaxies.  The
contour levels are chosen at 10, 30, 50, 70, and 90\% of the maximum value. 

\begin{figure}
\epsfxsize=8cm \epsfbox[-130 160 450 700]{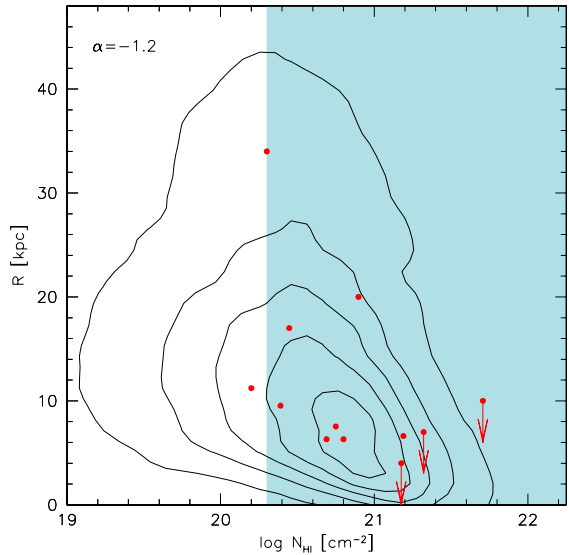} 
\caption{The two-dimensional probability distribution of cross section
in the $(\nhi,R)$ plane, where $R$ is distance from the centre of the
galaxy. The contours are from the \hi\ maps from the UMa sample, and the
dots represent data from \dla\ surveys. The shaded area indicates the
region that corresponds to \dla\ column densities.
\label{Pnew.fig}
}
\end{figure}                   

There are a number of interesting features in this plot.  First, it
shows that the largest concentration of \hi\ cross sections in galaxies
in the local universe is in column densities $20.6 < \log \nhi < 21.1$
and galactocentric radii $3<R<10\,\rm kpc$.  The observational fact that
the \hi\ distribution in galaxies often shows a central depression can
be seen in this figure by the compression of contours at $R\sim 0$.  Not
surprisingly, the outer contour shows that high column densities are
mostly seen at small $R$, but it also shows that column densities lower
than $\log\nhi=20.6$ are rarely seen in the extreme centres of galaxies.
(This might depend on the resolution).

In order to make a direct comparison between 21cm results and \dla\
parameters, we plotted on top of the contours the data points from \dla\
surveys taken from Le Brun et al.  (2000) and Nestor et al.  (2001). 
Note that these are the lowest redshift \dla\ systems known, but the
median redshift of the combined sample is $\langle z \rangle=0.5$, 
significantly higher
than the median $z=0.003$ for the UMa sample.  Evolutionary effects
might therefore be important, but at present a lower redshift \dla\
sample is not available. 

Although the statistics is still poor, there seems to be an agreement
between the distribution of the \dla\ points and the contours from the
21cm maps.  When more measurements of impact parameters and column
densities of low-$z$ \dla\ systems become available, this method will
provide a powerful tool to test whether the properties of \dla\ systems
and nearby galaxies are in agreement.


\section{Concluding remarks}
 We have shown that the properties of \dla\ galaxies are in good
agreement with the expectations based on 21cm images of nearby galaxies. 
Although most of the \hi\ {\em mass\/} density is locked up in luminous
spiral galaxies, the \hi\ {\em cross section\/} is much more spread out
over different galaxy properties.  Also the distribution of column
densities and impact parameters are in general agreement between \dla\
systems and local galaxies, although the statistics is still poor. 

There are a number of caveats that should be kept in mind when making
comparisons between QSO absorption line systems and galaxies.
First, the spatial resolution of the aperture synthesis instruments used
to derive \hi\ column density maps of nearby galaxies might have a
influence of the derived column density distribution functions. 
Beautiful examples of fine-structure in the \hi\ distribution in
galaxies can be found, for example, in the maps from the Southern
Galactic Plane Survey (McClure-Griffiths et al.  2001), in the Small
Magellanic Cloud maps of Stanimirovic et al.  (1999), or in Braun's (1995)
high resolution imaging of nearby spiral galaxies.
The results
presented here are based on much coarser resolution, which means that
most of this fine structure is averaged out.

A second point of concern is the small number of appropriately
studied galaxies in the available samples.  The cross
section distribution functions are derived from 49 galaxies, which
implies that there are on average 10 galaxies per bin in $M_B$, \mhi, or
surface brightness.  The calculation of cross-sectional areas
implicitly assumes that within each bin the galaxies are randomly
oriented, but for these small samples that might not always be
justified.  Also, it would be interesting to calculate the functions for
a larger range of galaxy parameters.
Both these issues will be addressed in a future paper in which we 
present an analysis of the column density distribution function of a much
larger galaxy sample taken from the WHISP survey (Westerbork
observations of neutral Hydrogen in Irregular and SPiral galaxies).

Finally, it is important to keep in mind the biases that are introduced
in assembling \dla\ galaxy samples.  As Turnshek et al.  (2001) stress,
the identification of a galaxy with the \dla\ absorber is more probable
if the impact parameter between the sight-line to the background QSO and
the galaxy is small.  On the other hand, if the impact parameter is very
small, the identification of a galaxy might be hindered by the
dominating QSO light.  \dla\ samples might be biased due to the effect
of dust obscuration in the \dla\ host galaxies.  Pei et al.  (1999) argue
that these `dusty sight-lines' might drop out of the QSO sample that is
used to identify the absorbing systems.


\end{document}